\documentclass[nonatbib]{ragtime} 

\title[Charged black holes embedded in organized magnetic fields]%
      {Astrophysical black holes embedded in organized magnetic fields~\\[5pt] 
      \it Case of a nonvanishing electric charge}

\author[V.~Karas]  
       {Vladim\'{\i}r Karas\\ 
        \ins{1}Astronomical~Institute, Czech Academy~of~Sciences, Bo\v{c}n\'{\i}~II~1401,\splitins[1]
        CZ-14100~Prague, Czech~Republic\\
        \ins{}E-mail:~\Email{vladimir.karas@asu.cas.cz} 
}
\begin{document}

\newcommand{\beq}{\begin{equation}}
\newcommand{\eeq}{\end{equation}}
\newcommand{\rd}{{\,\rm d}}
\newcommand{\ri}{{\rm{i}}}
\newcommand{\nab}{\mbox{\protect\boldmath$\nabla$}}
\renewcommand{\vec}[1]{\mbox{\protect\boldmath$#1$}}

\begin{abstract}
Large scale magnetic fields pervade the cosmic environment where the astrophysical black holes are often embedded and influenced by the mutual interaction. In this contribution we outline the appropriate mathematical framework to describe magnetized black holes within General Relativity and we  show several examples how these can be employed in the astrophysical context. In particular, we examine the magnetized black hole metric in terms of an exact solution of electro-vacuum Einstein-Maxwell equations under the influence of a non-vanishing electric charge. New effects emerge: the expulsion of the magnetic flux out of the black-hole horizon depends on the intensity of the imposed magnetic field.
\end{abstract}


\section{Introduction}
Astrophysical black holes are cosmic objects that can be mathematically described by a set of Einstein-Maxwell equations (e.g.\ Romero \& Vila 2014 \citep{rom14}). Various formulations of the Uniqueness Theorems express in a rigorous way the conditions under which the black hole solutions exist and they constrain the parameter space that is necessary to specify different cases (Wald 1984 \citep{wal84}). It turns out that classical black holes are described by a small number of such parameters, in particular, the mass, electric (or magnetic) charge, and angular momentum (spin). Black holes do not support their own magnetic field except the gravito-magnetically induced components in the rotating, charged Kerr-Newman metric.

However, astrophysical black holes are embedded in a magnetic field of external origin, which then interacts with the internal properties of the black hole (Ruffini \& Wilson \citep{ruf75}). In the case of very strong magnetic intensity, the magnetic field even contributes to the spacetime metric. In the present contribution we examine interesting properties of such an electrically charged, magnetized, rotating black hole. To this end we employ the solution originally derived in 1970s by means of Ernst magnetization techniques (Ernst \& Wild 1976 \citep{ern76}) and demonstrate its interesting features in terms of magnetic flux threading different regions of the black hole horizon or an entire hemisphere (see Bi\v{c}\'ak \& Hejda 2015 \citep{bic15}, and further references cited therein). 

We limit our discussion to axially symmetric and stationary solutions. These are vacuum, asymptotically non-flat solutions, where the influence of plasma is ignored but the effects of strong gravity are taken into account. 

\section{Magnetized black holes with spin and charge}
We can write the system of mutually coupled, Einstein-Maxwell partial differential equations (Chandrasekhar 1983 \citep{cha83}),
\beq 
R_{\mu\nu}-\textstyle{\frac{1}{2}}Rg_{\mu\nu}=8\pi T_{\mu\nu},
\eeq
where the source term $T_{\mu\nu}$ is of purely electromagnetic origin,
\beq
T^{\alpha\beta}\equiv T^{\alpha\beta}_{\rm EMG}=\frac{1}{4\pi}\left(F^{\alpha\mu}F^\beta_\mu- \frac{1}{4}F^{\mu\nu}F_{\mu\nu}g^{\alpha\beta}\right),
\eeq
\beq
{T^{\mu\nu}}_{;\nu}=-F^{\mu\alpha}j_{\alpha},\qquad
{F^{\mu\nu}}_{;\nu}=4\pi j^\mu,\qquad {^\star F^{\mu\nu}}_{;\nu}=4\pi\mathcal{M}^\mu,
\eeq
and $^\star F_{\mu\nu}\equiv\frac{1}{2}{\varepsilon_{\mu\nu}}^{\rho\sigma}F_{\rho\sigma}$.
We will consider the spacetime solutions for the metric that satisfies electro-vacuum case with a regular event horizon under the constraints of axial symmetry and stationarity,
\beq
\rd s^2=f^{-1}\left[e^{2\gamma}\left(\rd z^2+\rd\rho^2\right)+\rho^2\rd\phi^2\right]-f\left(\rd t-\omega\rd\phi\right)^2,
\label{ds2}
\eeq
with $f$, $\omega$, and $\gamma$ being functions of $z$ and $\rho$ only. Hereafter, instead of the canonical cylindrical form (\ref{ds2}), we will employ also the spheroidal coordinates $r$ and $\theta$ when convenient. In the weak electromagnetic field approximation, the electromagnetic (test) field is supposed to reside in the background of a rotating black hole, e.g., Kerr metric or a weakly charged Kerr metric (e.g. Wald 1984 \citep{wal84}, Gal'tsov 1986 \citep{gal86}). 
As an example, in such an asymptotically flat spacetime, the axial Killing vector $\partial_\phi(\equiv\tilde{\xi})$ generates a uniform magnetic field, whereas the field vanishes asymptotically for the time-like Killing vector $\partial_t(\equiv\xi)$. These two solutions are known as the Wald's field, i.e.
\beq
\mbox{\protect\boldmath$F$}=\textstyle{\frac{1}{2}}B_0\left(\mbox{\protect\boldmath$\rd$}\tilde{\xi}+\frac{2J}{M}\mbox{\protect\boldmath$\rd$}\xi\right)
\eeq
in ref.\ \citet{wal74} notation. Magnetic flux surfaces are defined,
\beq
4\pi \Phi_{{\rm m}}=\int_\mathcal{S}\mbox{\protect\boldmath$F$}\;=\;\mbox{const}.
\eeq
Magnetic and electric (Lorentz) forces are then given in terms of particle's mass $m$, its four-velocity \mbox{\protect\boldmath$u$}, and magnetic and electric charges,$q_{\rm m}$ and $q_{\rm e}$, respectively:
\beq
m\mbox{\boldmath$\dot{u}$}=q_{\rm{m}}\mbox{\boldmath${^\star}F.u$},\qquad m\mbox{\boldmath$\dot{u}$}=q_{\rm{e}}\mbox{\boldmath$F.u$}.
\eeq
Magnetic field lines are determined by
\beq
\frac{{\rd{r}}}{{\rd\theta}}=\frac{B_r}{B_\theta},\qquad \frac{{\rd{r}}}{{\rd\phi}}=\frac{B_r}{B_\phi},
\label{l1}
\eeq
in a perfect analogy with classical electromagnetism. We will employ the above-given quantities in our discussion further below.

Magnetic (electric) lines of force are defined by the direction of Lorentz force that acts on electric (magnetic) charges,
\beq
\frac{\rd u^\mu}{\rd\tau}\propto \,^{\star}\!F^\mu_\nu\,u^\nu,\qquad \frac{\rd u^\mu}{\rd\tau}\propto F^\mu_\nu\,u^\nu.
\eeq
Therefore, in an axially symmetric system, the equation for magnetic lines takes a lucid form,
\beq
\frac{\rd r}{\rd\theta}=-\frac{F_{\theta\phi}}{F_{r\phi}},\qquad \frac{\rd r}{\rd\phi}=\frac{F_{\theta\phi}}{F_{r\theta}},
\eeq
that is in correspondence with eq.~(\ref{l1}).

Let us now turn our attention to the case of strong magnetic field, where we cannot ignore its influence on the spacetime metric. The latter is not necessarilly flat in the asymptotical spatial region far from the black hole (Ernst \& Wild 1976 \citep{ern76}; Karas \& Vokrouhlick\'y 1990 \citep{kar90}). 

As an example, we can consider magnetized Kerr-Newman black hole metric expressed in the form (Garc\'{\i}a D\'{\i}az 1985 \citep{gar85})
\begin{eqnarray}
\rd s^2&\,=\,&|\Lambda|^2\Sigma\left(\Delta^{-1}\rd{r}^2+\rd{\theta}^2-\Delta{A^{-1}}\rd{t}^2\right) \nonumber \\[3pt]
&&+|\Lambda|^{-2}\Sigma^{-1}A\sin^2\theta\left(\rd{\phi}-\omega\rd{t}\right)^2,
\label{g}
\end{eqnarray}
where $\Sigma=r^2+a^2\cos^2\theta$, $\Delta=r^2-2Mr+a^2+e^2$, $A=(r^2+a^2)^2-{\Delta}a^2\sin^2\theta$ are functions from the Kerr-Newman metric. The outer horizon is located at radius $r{\equiv}r_+=1+(1-a^2-e^2)^{1/2}$, like in an unmagnetized case, and the horizon existence is restricted to the range of parameters $a^2+e^2\leq1$. Let us emphasise that, in the magnetized case, the traditional Kerr-Newman parameters $a$ and $e$ are {\em not identical} with the black hole total spin and electric charge, as we will see further below. Moreover, because of asymptotically non-flat nature of the spacetime, the Komar-type angular momentum and electric charge (as well as the black hole mass) have to be defined by integration over the horizon sphere rather than at radial infinity.

The magnetization function $\Lambda=1+\beta\Phi-\frac{1}{4}\beta^2\mathcal{E}$, in terms of the Ernst potentials $\Phi(r,\theta)$ and $\mathcal{E}(r,\theta)$, reads
\begin{eqnarray}
\Sigma\Phi
 &=& ear\sin^2\theta-{\ri}e\left(r^2+a^2\right)\cos\theta, \\
\Sigma\mathcal{E}
 &=& -A\sin^2\theta-e^2\left(a^2+r^2\cos^2\theta\right)
 \nonumber \nonumber \\
 & & + 2{\ri}a\left[\Sigma\left(3-\cos^2\theta\right)+a^2\sin^4\theta-
 re^2\sin^2\theta\right]\cos\theta.
\end{eqnarray}

\begin{figure}
\begin{center}
\includegraphics[width=0.8\textwidth]{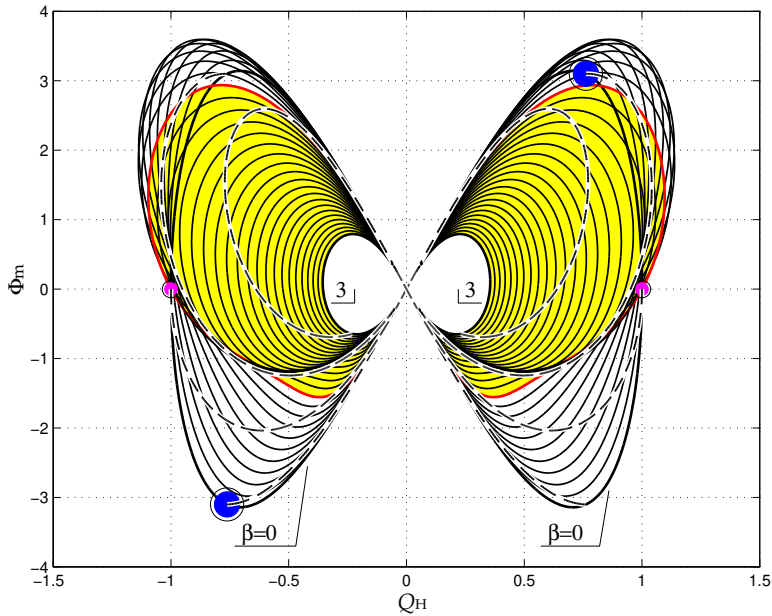}
\end{center}
\caption{The ``butterfly diagram'' shows the magnetic flux $\Phi_{\rm m}$ of magnetized Kerr-Newman black hole with $a^2+e^2=1$ as a function of the total electric charge $Q_{\rm H}$. Solid curves correspond to a constant value of the dimensionless magnetization parameter $\beta=BM$ ($\beta=0$ is the case of an unmagnetized Kerr-Newman black hole). The area of the plot with ultra-strong magnetization is bounded by $\beta=1$ (red curve) and emphasized by yellow colour in the plot. The lines of constant ratio of $a/e$ and varying $\beta$ are also plotted (dashed; the cases of $a/e=\pm0.85$ and $0$ are shown); some distinctive combinations of the parameters $a$, $e$ are emphasized by colour points.}
\label{fig1}
\end{figure}

The corresponding components of the electromagnetic field can be written conveniently with respect to orthonormal LNRF components,
\begin{eqnarray}
H_{(r)}+{\ri}E_{(r)} &=& A^{-1/2}\sin^{-1}\!\theta\,\Phi^{\prime}_{,\theta},\\
H_{(\theta)}+{\ri}E_{(\theta)} &=&-\left(\Delta/A\right)^{1/2}\sin^{-1}\!\theta\,\Phi^{\prime}_{,r},
\label{heth}
\end{eqnarray}
where $\Phi^{\prime}(r,\theta)=\Lambda^{-1}\left(\Phi-\frac{1}{2}\beta\mathcal{E}\right)$.
The total electric charge $Q_{\rm{H}}$ is
\begin{equation}
Q_{\rm{H}} = -|\Lambda_0|^2\,\Im{\rm{m}\,}\Phi^{\prime}\left(r_+,0\right),
\label{qh} 
\end{equation}
and the magnetic flux $\Phi_{\rm{m}}(\theta)$ across a cap placed in an axisymmetric position on the horizon is
\begin{equation}
\Phi_{\rm{m}} = 2\pi|\Lambda_0|^2\,\Re{\rm{e}\,}\Phi^{\prime}
 \left(r_+,\bar{\theta}\right)\Bigr|\strut^{\theta}_{\bar{\theta}=0},
\label{fh}
\end{equation}
where $\Lambda_0=\Lambda(\theta=0)$. The adopted notation indicates subtraction of the Ernst potential values $\Phi^\prime\left(r_+,\bar{\theta}\right)$ taken at $\bar{\theta}\rightarrow\theta$ and $\bar{\theta}\rightarrow0$.

At this point it is interesting to mention that the span of the azimuthal coordinate in the magnetized solution must be rescaled by the multiplication factor $\Lambda_0$ in order to avoid a conical singularity on the symmetry axis (Hiscock 1981 \citep{his81}):
\begin{equation}
\Lambda_0 = \left[1+\textstyle{\frac{3}{2}}\beta^2e^2+2\beta^3ae+\beta^4\left(\textstyle{\frac{1}{16}}e^4+a^2\right)\right]^{1/2}.
\end{equation}
This rescaling procedure effectively leads to the increase of the horizon surface area, and thereby also magnetic flux across the horizon (Karas 1988 \citep{kar88}). In Figure \ref{fig1}, the magnetic flux across the entire black hole hemisphere in Kerr-Newman strongly magnetized black hole solution, $\Phi_{\rm{m}}(\theta=\pi/2)$, is shown as a function of electric charge on the horizon, $Q_{\rm{H}}$ (additional details can be found in Karas et al.\ \citep{kar00,kar21}). 

Cases of intersection of the $\beta={\rm const}$ curves with $F=0$ and non-zero charge, $Q_{\rm{H}}\neq0$, correspond to the vanishing angular momentum of the black hole, $J=0$. This property is rather different from the behaviour of weakly magnetized black holes with only test magnetic field imposed on them. On the other hand, this exact solution does not allow us to study the effects of mis-alignment of the magnetic field with respect to the rotation axis, which is so far possible only in the test-field approximation or by numerical techniques. Let us also note that it may be interesting to consider the magnetic flux also in other spacetime metrics for comparison and better understanding of the underlying processes (see, e.g., Gutsunaev et al.\ 1988 \citep{gut88}, Kov\'a\v{r} et al.\ 2013 \citep{kov13}, Khan \& Chen 2023 \citep{kha23}, Vrba et al. 2023 \citep{vrb23}, and further references cited therein).

\section{Conclusions}
We discussed the magnetic flux across the event horizon of a magnetized rotating black hole and the associated electric charge within the framework of the {\em exact solution} of mutually coupled Einstein-Maxwell equations. To this end, we adopted Ernst-Wild \citep{ern76} spacetime, which represents an axially symmetric, stationary solution that corresponds, in a straightforward manner, to the Wald \citep{wal74} solution of an asymptotically uniform magnetic field imposed on the background of Kerr metric. On the other hand, in the limit of small black hole mass, zero angular momentum and strong magnetic intensity, the adopted solution goes over to the cosmological solution of Melvin universe \citep{mel64}. The latter represents an asymptotically non-flat `geon' that is maintained in the static configuration by gravitational effect of the magnetic field acting upon itself. In other words, this means that our discussion is appropriate for the limit of ultra-strong magnetic fields that might be possibly relevant in the conditions of early Universe \citep{bes15}. On the other hand, we think that this extreme situation does not bring any qualitatively new phenomena into the discussion of the magnetic Penrose process, which has been recently discussed in the context of weakly magnetized black holes \citep{kar23}.

We elaborated a detailed graphical representation of magnetic flux which exhibits an intricate structure as a function of the space-time parameters. This has allowed us to position specific configurations, such as those with vanishing angular momentum or vanishing electric charge. The corresponding combinations of the model parameters are different from the case of weak magnetic field limit, because of strong-gravity effects. Let us note that ultra-strong magnetic fields are expected to affect processes on molecular and atomic scales, in the conditions when the magnetic energy density becomes comparable to the binding energy density \citep{lai15}.


\ack
The author thanks the referee for careful reading of the original version of the manuscript and helpful suggestions towards improvements at several points. The author acknowledges continued support from the Czech Science Foundation EXPRO grant titled ``Accreting black holes in the new era of X-ray polarimetry missions'', No.\ 21-06825X.

\end{document}